\documentclass[12pt]{article}
\usepackage{epsfig,amsfonts,amssymb}
\begin{document}

\voffset -0.7 true cm
\hoffset 1.1 true cm
\topmargin 0.0in
\evensidemargin 0.0in
\oddsidemargin 0.0in
\textheight 8.6in
\textwidth 7.1in
\parskip 10 pt

\newcommand{\be}{\begin{equation}}
\newcommand{\ee}{\end{equation}}
\newcommand{\bea}{\begin{eqnarray}}
\newcommand{\eea}{\end{eqnarray}}
\newcommand{\beas}{\begin{eqnarray*}}
\newcommand{\eeas}{\end{eqnarray*}}

\def\kl{{\frac{2 \pi l}{\beta}}}
\def\km{{\frac{2 \pi m}{\beta}}}
\def\kn{{\frac{2 \pi n}{\beta}}}
\def\kr{{\frac{2 \pi r}{\beta}}}
\def\ks{{\frac{2 \pi s}{\beta}}}
\def\b{{\beta}}
\font\cmsss=cmss8
\def\C{{\hbox{\cmsss C}}}
\font\cmss=cmss10
\def\bigC{{\hbox{\cmss C}}}
\def\scriptlap{{\kern1pt\vbox{\hrule height 0.8pt\hbox{\vrule width 0.8pt
  \hskip2pt\vbox{\vskip 4pt}\hskip 2pt\vrule width 0.4pt}\hrule height 0.4pt}
  \kern1pt}}
\def\ba{{\bar{a}}}
\def\bb{{\bar{b}}}
\def\bc{{\bar{c}}}

\def\Bigggl{\mathopen\Biggg}
\def\Bigggr{\mathclose\Biggg}
\def\Biggg#1{{\hbox{$\left#1\vbox to 25pt{}\right.\n@space$}}}
\def\n@space{\nulldelimiterspace=0pt \m@th}
\def\m@th{\mathsurround = 0pt}

\begin{titlepage}
\begin{flushright}
{\small BROWN-HET-1276} \\
{\small CU-TP-1025} \\
{\small hep-th/0108006}
\end{flushright}

\begin{center}

\vspace{2mm}

{\Large \bf Probing Black Holes in Non-Perturbative Gauge Theory}

\vspace{3mm}

Norihiro Iizuka${}^1$, Daniel Kabat${}^1$, Gilad Lifschytz${}^2$ and David A.\ Lowe${}^3$

\vspace{1mm}

${}^1${\small \sl Department of Physics} \\
{\small \sl Columbia University, New York, NY 10027} \\
{\small \tt iizuka, kabat@phys.columbia.edu}
\vspace{1mm}

${}^2${\small \sl Department of Mathematics and Physics} \\
{\small \sl University of Haifa at Oranim, Tivon 36006, ISRAEL} 
{\small \tt giladl@research.haifa.ac.il}

${}^3${\small \sl Department of Physics} \\
{\small \sl Brown University, Providence, RI 02912} \\
{\small \tt lowe@het.brown.edu}

\end{center}

\vskip 0.3 cm

\noindent
We use a 0-brane to probe a ten-dimensional near-extremal black hole
with $N$ units of 0-brane charge.  We work directly in the dual
strongly-coupled quantum mechanics, using mean-field methods to
describe the black hole background non-perturbatively.  We obtain the
distribution of W boson masses, and find a clear separation between
light and heavy degrees of freedom.  To localize the probe we
introduce a resolving time and integrate out the heavy modes.  After a
non-trivial change of coordinates, the effective potential for the
probe agrees with supergravity expectations.  We compute the entropy
of the probe, and find that the stretched horizon of the black hole
arises dynamically in the quantum mechanics, as thermal restoration of
unbroken $U(N+1)$ gauge symmetry.  Our analysis of the quantum
mechanics predicts a correct relation between the horizon radius and
entropy of a black hole.

\end{titlepage}

%%%%%%%%%%%%%%%%%%%%%%%%%%%%%%%%%%%%%%%%%%%%%%%%%%%%%%%%%%%%%%%%%%%%%%%%%%%%%%
\section{Introduction}
%%%%%%%%%%%%%%%%%%%%%%%%%%%%%%%%%%%%%%%%%%%%%%%%%%%%%%%%%%%%%%%%%%%%%%%%%%%%%%

For many years it has been an outstanding challenge to develop a
microscopic understanding of black hole physics.  Many properties of
black holes can be easily understood using classical or semiclassical
gravity.  For example the notion of a horizon arises in classical
gravity, while semiclassical considerations show that a horizon has an
associated thermodynamic entropy.  But ultimately we must understand
how these semiclassical properties arise from a microscopic theory of
quantum gravity.

The development of non-perturbative definitions of string theory
\cite{reviews} has given a new perspective on these problems.  In
particular string theory in the background of a ten-dimensional
non-extremal black hole with $N$ units of 0-brane charge is known to
have a dual description in terms of $U(N)$ gauged supersymmetric
quantum mechanics with sixteen supercharges \cite{imsy}.  We would
like to understand how the semiclassical physics of black holes
emerges from the dual quantum mechanics.  Can we recover the classical
geometry of the black hole?  Can we understand the horizon in terms of
microscopic physics?  Can we account for the Hawking-Bekenstein
entropy?

These questions are not easily answered, because the dual quantum
mechanics is strongly coupled whenever semiclassical supergravity is
valid \cite{imsy}.  In some cases, one can rely on supersymmetric
non-renormalization theorems to calculate at strong coupling
\cite{NonRenorm}.  But to make progress in a more general
setting, we need non-perturbative methods to study the quantum
mechanics.

In \cite{KabatLifschytz,KLL} we developed a mean-field approximation
scheme for the quantum mechanics of $N$ D0-branes.  Similar techniques
have been applied to matrix integrals in \cite{OdaSugino}.  The
approximation can be applied at strong coupling, and gives results for
thermodynamic quantities which are in good agreement with
semiclassical black hole predictions, at least over a certain range of
temperatures.  This supports the claim that the dual quantum mechanics
can account for the Hawking-Bekenstein entropy.  But to address the
other questions listed above, we need a probe that is sensitive to the
geometry of the black hole.

In the present paper, we introduce an additional D0-brane as a probe of
the black hole geometry.  It is easy to describe the probe in terms of
classical supergravity.  In the dual quantum mechanics, the description is in
terms of a gauge theory spontaneously broken to $U(N) \times U(1)$.  We
will make use of our mean-field approximation to describe the quantum
mechanics of the $U(N)$ black hole background.  There are a number of
interesting questions that we can address in this setting.  Can one
describe a localized probe in the quantum mechanics?  Can one recover
the expected effective potential for the probe?  What physics is
responsible for the horizon of the black hole?

An outline of this paper is as follows.  In section 2 we review the
supergravity description of a D0-brane probe of the black hole
background.  In section 3 we apply mean-field methods to the dual
quantum mechanics problem, and show that we can localize the probe by
introducing a resolving time in the quantum mechanics.  In section 4
we perform a spectral analysis of 2-point functions in the quantum
mechanics, to obtain the microscopic density of single-string
excitations.  We show that light states are present at the horizon;
thus the horizon can be understood as thermal restoration of unbroken
$U(N+1)$ gauge symmetry in the quantum mechanics.  In section 5 we
show that the effective potential for the probe agrees with
supergravity expectations.  The microcanonical entropy of the probe is
computed in section 6.  In section 7 we show that our results imply a
correct relation between the horizon radius and entropy of a black
hole.  Section 8 gives some conclusions and directions for future
work.

%%%%%%%%%%%%%%%%%%%%%%%%%%%%%%%%%%%%%%%%%%%%%%%%%%%%%%%%%%%%%%%%%%%%%%%%%%%%%%
\section{Supergravity predictions}
%%%%%%%%%%%%%%%%%%%%%%%%%%%%%%%%%%%%%%%%%%%%%%%%%%%%%%%%%%%%%%%%%%%%%%%%%%%%%%

We begin with a review of the supergravity description of the probe /
black hole system.  We will be able to extract a number of useful
predictions about the behavior of the dual quantum mechanics.

Our focus will be on the near-horizon region of a ten-dimensional
non-extremal black hole in type IIA supergravity.  The black hole is
taken to have $N$ units of D0-brane charge, so that the dual quantum
mechanics is $U(N)$ gauged supersymmetric quantum mechanics with
sixteen supercharges \cite{imsy}.  In the near-horizon region the
string-frame metric of the black hole is given by
\bea
\nonumber
ds^2 & = & \alpha'\left[-h(U)dt^2 + h^{-1}(U)dU^2 + {c^{1/2}
(g_{YM}^2 N)^{1/2} \over U^{3/2}} d\Omega_{8}^{2}\right] \\
\label{metric}
h(U) & = & \frac{U^{7/2}}{c^{1/2} (g_{YM}^2 N)^{1/2}}\left(1-\frac{U_{0}^{7}}{U^{7}}\right)
\eea
where $c = 2^7 \pi^{9/2} \Gamma(7/2)$ and
$g^2_{YM} = g_s / 4 \pi^2 (\alpha')^{3/2}$ is the coupling constant of
the dual gauge theory.  The dilaton profile is given by
\be
\label{dilaton}
e^\phi = {1 \over (\alpha')^{3/2}} \left({c g^2_{YM} N \over U^7}\right)^{3/4}
\ee
and there is a R-R 1-form potential \cite{Kiritsis}
\be
\label{RR}
A_0 = 1 + \frac{(\alpha')^2 U^7}{c g^2_{YM} N}\left(-1 + {U_0^7 \over 2 U^7}\right) + {\cal O}(\alpha'{}^4)\,.
\ee

The horizon of the black hole is located at $U=U_0$, which corresponds to a
Hawking temperature
\be
\label{HawkingTemp}
T = {7\over 2\pi \sqrt{30}} ~ (g^2_{YM} N)^{-1/2} \left(U_0 \over 2 \pi \right)^{5/2}
= 0.2034 ~ (g^2_{YM} N)^{-1/2} \left(U_0 \over 2 \pi \right)^{5/2}\,.
\ee
This is the temperature measured with respect to the Schwarzschild time coordinate
$t$.  Since $t$ is identified with the time coordinate of the dual
gauge theory, the dual quantum mechanics is to be studied at the same
finite temperature.  The black hole has a free energy
\cite{KlebanovTseytlin}
\be
\label{beken}
\beta F = - \left( { 2^{21} 3^2 5^7 \pi^{14} \over
7^{19}}\right)^{1/5} \!\! N^2 \left({T \over (g_{YM}^2 N)^{1/3}}
\right)^{9/5} = - 4.115 ~ N^2 \left({T \over (g_{YM}^2 N)^{1/3}} \right)^{1.8}\,.
\ee
Duality predicts that the quantum mechanics should have the same free
energy.  The supergravity description is expected to be valid when the
curvature and the dilaton are small near the black hole horizon.  This
regime corresponds to the 't Hooft large-$N$ limit of the quantum
mechanics, together with the requirement that the effective 't Hooft
coupling in the quantum mechanics
\be
\label{EffectiveCoupling}
g_{\rm eff}^2 = g^2_{YM} N / T^3
\ee
lies in the range
\be
\label{CouplingRange}
1 \ll g_{\rm eff}^2 \ll N^{10/7}\,.
\ee
Note that the quantum mechanics is strongly coupled whenever semiclassical
supergravity is valid \cite{imsy}.

A 0-brane probe of this supergravity background is described by the action
\be
\label{DBI}
S = - T_0 \int dt \, e^{-\phi} \sqrt{- \det G} - T_0 \int dt A_0
\ee
where the tension of a 0-brane is $T_0 = 1/g_s \sqrt{\alpha'}$.
Evaluating this on the black hole background
(\ref{metric}) -- (\ref{RR}) gives the effective action for the probe
in the decoupling limit \cite{ProbeAction,Kiritsis}
\bea
\label{probeDBI}
S & = & - {1 \over 4 \pi^2 g^2_{YM}} \int dt\, \Bigggl[ {U^7 \over c g^2_{YM} N}\left(- 1 + {U_0^7 \over 2 U^7}\right) \\
\nonumber
& & + \left({U^7 \over c g^2_{YM} N}\right)^{3/4} \sqrt{h(U)
- {1 \over h(U)} \dot{U}^2 - {c^{1/2} (g^2_{YM} N)^{1/2} \over U^{3/2}} \dot{\Omega}^2} \Bigggr]
\eea
(we dropped a constant term, the rest energy of the probe at
infinity).  From the action we can read off the effective potential
for a static probe,
\be
\label{SugraPotential}
V_{\rm eff} = -  \frac{N}{15(g_{YM}^2 N)^2}\left({U \over 2 \pi}\right)^7 \left(\sqrt{1 - {U_0^7 \over U^7}} - 1\right)^2\,.
\ee

Note that the effective potential (\ref{SugraPotential}) 
is singular at the horizon of the
black hole.  A singularity in a low-energy effective action suggests
that massless degrees of freedom have been improperly integrated out.
To see that this is indeed the case, consider `W-bosons': open strings
which can stretch between the probe 0-brane and the black hole.  The
energy of these strings can be computed by evaluating the Nambu-Goto
action
\[
S_{NG} = - {1 \over 2 \pi \alpha'} \int d^2 \sigma \sqrt{- \det G}
\]
on a string worldsheet which starts at the horizon and ends on the
probe.  A straightforward calculation gives an energy, measured with
respect to the Schwarzschild time coordinate $t$, equal to
\be
\label{mW}
m_W = {1 \over 2 \pi} \left(U - U_0\right)\,.
\ee
This is identical to the calculations of \cite{Wilson}, performed in the
context of studying Wilson lines at finite temperature.

The energy of these strings indeed goes to zero as the probe
approaches the horizon.  In \cite{causality} it was argued that these
massless degrees of freedom are responsible for the singularity of the
supergravity effective potential (\ref{SugraPotential}).  We conclude
that the supergravity description of the probe breaks down when the
probe gets too close to the horizon, at least according to a
Schwarzschild observer.\footnote{It is an outstanding question to
understand the horizon microscopically from the point of view of a
freely falling observer.}  In the full string theory (or dual quantum
mechanics), which takes these light degrees of freedom into account,
the W-bosons will become thermally excited as the probe approaches the
horizon.

To estimate the radius of the `stretched horizon' at which this
thermalization starts to occur, we compare the energy of a W-boson (\ref{mW})
to the temperature of the black hole (\ref{HawkingTemp}) (the comparison is meaningful,
since both energies are measured with respect to the same
Schwarzschild time coordinate $t$).  Setting $m_W = T$ gives the
radius of the stretched horizon
\be
\label{Us}
U_{\rm stretched} = U_0 \left(1 + {7 \over \sqrt{30} (2 \pi)^{5/2}}
\biggl({U_0 \over (g^2_{YM} N)^{1/3}}\biggr)^{3/2}\right)\,.
\ee
At this radius stringy degrees of freedom start to become thermally
excited, so the supergravity description of the probe breaks down, and
the classical black hole geometry (\ref{metric}) is no longer reliable.

For an accurate description of physics inside the stretched horizon, we
must turn to the dual quantum mechanics (or full string theory).
There we see that light W-bosons are rapidly created, so the probe
quickly thermalizes with the black hole.  Once the probe 0-brane has
come to thermal equilibrium, it can no longer be distinguished from
the other 0-branes that make up the black hole.

This makes it easy to compute the energy and entropy of the probe once
it has been absorbed by the black hole and come to equilibrium.  The
effect of the probe is simply to shift $N \rightarrow N+1$ in the
semiclassical free energy of the black hole (\ref{beken}).  We shift
$N$ holding both $g_{YM}$ and the temperature fixed.  Thus the free
energy of a probe in equilibrium is given by
\be
\label{shiftun}
\beta F_{\rm probe} = - 5.76 N \left(T \over (g^2_{YM} N)^{1/3}\right)^{9/5}\,.
\ee
This agrees precisely with the free energy one obtains by evaluating
the DBI effective potential (\ref{SugraPotential}) at the horizon of
the black hole, $\beta F_{\rm probe} = \beta V_{\rm eff}
\vert_{U=U_0}$.  This rather surprising coincidence was noted and
explained in \cite{KiritsisTaylor}.

To better understand the absorption process, note that the energy and
entropy of a probe in equilibrium can be obtained from the
corresponding black hole quantities by shifting $N \rightarrow N+1$.
That is, they are given by
\bea
\nonumber
E_{\rm probe} & = & \left( { 2^{21} 3^{12} \pi^{14} \over
5^3 7^{14} }\right)^{1/5} \!\! N  (g_{YM}^2 N)^{1/3}\left({T \over (g_{YM}^2 N)^{1/3}}
\right)^{14/5} \\
S_{\rm probe} & = & \left( { 2^{26} 3^2  \pi^{14} \over
5^3 7^{9}}\right)^{1/5} \!\! N \left({T \over (g_{YM}^2 N)^{1/3}}
\right)^{9/5} \!\!\! .
\eea
Note that the probe acquires positive energy as it comes to
equilibrium with the black hole.  This is in sharp contrast to the
negative attractive potential (\ref{SugraPotential}) which is present
outside the stretched horizon.  Moreover the probe acquires entropy
once it is inside the stretched horizon, associated with the excited
W-boson degrees of freedom.  By contrast, outside the stretched
horizon the probe entropy is very small (according to classical
supergravity, it is exactly zero).  Thus absorption of the probe by the
black hole is driven by the increase in entropy, which more than makes
up for the cost in energy.

%%%%%%%%%%%%%%%%%%%%%%%%%%%%%%%%%%%%%%%%%%%%%%%%%%%%%%%%%%%%%%%%%%%%%%%%%%%%%%
\section{Gauge theory calculations}
%%%%%%%%%%%%%%%%%%%%%%%%%%%%%%%%%%%%%%%%%%%%%%%%%%%%%%%%%%%%%%%%%%%%%%%%%%%%%%

We now turn to the dual gauge theory description of the probe / black
hole system, and apply mean-field methods \cite{KLL} to study the
quantum mechanics at strong coupling.

Our starting point is $U(N+1)$ gauged supersymmetric quantum mechanics
with sixteen supercharges \cite{ClaudsonHalpern}.  As in our previous
work, we will describe the quantum mechanics using the language of
${\cal N} = 2$ superspace.  This formalism makes ${\cal N} = 2$
supersymmetry manifest (out of the underlying ${\cal N} = 16$).  It
also makes manifest a $G_2 \times SO(2)$ subgroup of the underlying
$SO(9)$ rotational invariance.

In terms of ${\cal N} = 2$ superfields, the action for 0-brane quantum
mechanics reads
\be
\label{action}
S = {1 \over g^2_{YM}} \int dt d^2\theta \, {\rm Tr} \biggl(
- {1 \over 4} \nabla^\alpha {\cal F}_i \nabla_\alpha {\cal F}_i
- \frac{1}{2} \nabla^\alpha \Phi_A \nabla_\alpha \Phi_A
- \frac{i}{3} f_{ABC} \Phi_A [\Phi_B, \Phi_C] \biggr)\,.
\ee
Here $\nabla_\alpha$ is a $U(N+1)$ gauge superconnection, and ${\cal
F}_i$ is the corresponding field strength.  The fields
\[
\Phi_A = \phi_A + i \psi_{A \alpha} \theta_\alpha + f_A \theta^2
\]
are a collection of seven adjoint scalar multiplets, with $A,\, B =
1,\ldots,7$ an index in the ${\bf 7}$ of $G_2$, and $f_{ABC}$ is a
suitably normalized totally antisymmetric $G_2$-invariant tensor.  For
more details on our notation see appendix A.

%%%%%%%%%%%%%%%%%%%%%%%%%%%%%%%%%%%%%%%%%%%%%%%%%%%%%%%%%%%%%%%%%%%%%%%%%%%%%%
\subsection{Introducing a localized probe}
%%%%%%%%%%%%%%%%%%%%%%%%%%%%%%%%%%%%%%%%%%%%%%%%%%%%%%%%%%%%%%%%%%%%%%%%%%%%%%

We wish to use a single 0-brane to probe a black hole with $N$ units
of 0-brane charge.  To do this we separate the probe from the black
hole, by giving an expectation value to one of the scalar fields, say
\be
\label{vev}
\langle \Phi_7 \rangle = \left(\begin{array}{cc}
0 & 0 \\ 0 & R \end{array}\right)\,.
\ee
Classically this breaks the gauge symmetry from $U(N+1)$ down to $U(N)
\times U(1)$, and gives the off-diagonal fields a mass $m_W = R$.

Before continuing, there is an important question we must address: is
the expectation value (\ref{vev}) meaningful at strong coupling?  The
issue is that at strong coupling the eigenvalues of $\phi_A$ have
large quantum fluctuations.  These fluctuations can be measured by
computing the connected equal-time correlation function ${1 \over N}
\langle {\rm Tr} (\phi_A)^2 \rangle_C$.  For example in our
approximation the eigenvalues obey a Wigner semicircle distribution,
with maximum eigenvalue
\be
\label{Wigner}
\lambda_{\rm max} = 2 \sqrt{\langle {\rm Tr} (\phi_A)^2 \rangle_C / N}\,.
\ee
On general grounds one can argue that \cite{Polchinski,Susskind}
\be
\label{fluctuations}
{1 \over N} \langle \, {\rm Tr} (\phi_A)^2 \, \rangle_C \, \sim \, \left(g^2_{YM} N\right)^{2/3}\,,
\ee
and this behavior is indeed seen in our mean-field approximation
\cite{KLL}.  Thus the eigenvalues of $\phi_A$ fluctuate
over the entire region, of size $\sim (g^2_{YM} N)^{1/3}$ \cite{imsy},
in which supergravity is valid.  We want to place the probe well
inside the supergravity region, at some value of $R \ll (g^2_{YM}
N)^{1/3}$.  But can we really claim to have a well-localized probe,
given the large fluctuations (\ref{fluctuations})?

The answer is that the probe can be well-localized, as long as it is
outside the stretched horizon.  To see this, note that supergravity
only emerges as a low energy approximation to the quantum mechanics.
To discuss the position of the probe we must introduce a resolving
time, and integrate out the high-frequency degrees of freedom in the
quantum mechanics\footnote{We are indebted to Lenny Susskind for
emphasizing this point to us on numerous occasions.  We are also
grateful to Emil Martinec for valuable discussions on this topic.}.
These high-frequency modes are responsible for the large fluctuations
(\ref{fluctuations}), and by integrating them out, we can construct a
sharply-defined position operator for the probe.

Following \cite{Susskind}, a suitable position operator can be
obtained by smearing the Heisenberg picture fields over a Lorentzian
time interval $\epsilon$.
\be
\label{SmearedOperator}
\bar{\phi}_A = \int_{-\infty}^\infty {dt \over \epsilon \sqrt{\pi}} e^{-t^2/\epsilon^2} \phi_A(t)
\ee
The effect of the resolving time $\epsilon$ is to integrate out modes
with frequency larger than $1/\epsilon$.  We can only integrate out
modes which are not thermally excited, so an appropriate choice of
resolving time is $\epsilon \sim \beta$.  Then the operators
$\bar{\phi}_A$ provide sharply-defined position operators for the
probe, at least as long as the probe is outside the stretched horizon.
As we shall see, as the probe enters the stretched horizon,
W-bosons with a mass of order $1/\beta$ start to become thermally
excited.  These light W-bosons contribute to the fluctuations in
$\bar{\phi}_A$, so once the probe enters the stretched horizon it cannot be localized to less than the size of the stretched
horizon.

Although the probe can be well-localized outside the stretched
horizon, one subtle question remains: what is the precise relationship
between the Higgs vev $R$ appearing in (\ref{vev}) and the
supergravity radial coordinate $U$ appearing in (\ref{metric})?  At
zero temperature one can rely on supersymmetry to make a precise
identification.  The mass of a BPS stretched string in the gauge
theory is given exactly by the tree-level formula $m_W = R$, while in
supergravity the classical formula $m_W = U / 2 \pi$ (\ref{mW}) is not
corrected.  Therefore one can identify $R = U / 2 \pi$ at zero
temperature.  However, this identification is not
appropriate at finite temperature.  We will work out the correct
identification in section 5.

%%%%%%%%%%%%%%%%%%%%%%%%%%%%%%%%%%%%%%%%%%%%%%%%%%%%%%%%%%%%%%%%%%%%%%%%
\subsection{Mean-field approximation}
%%%%%%%%%%%%%%%%%%%%%%%%%%%%%%%%%%%%%%%%%%%%%%%%%%%%%%%%%%%%%%%%%%%%%%%%

Having understood the description of a localized probe at strong
coupling, we proceed to apply mean-field methods \cite{KabatLifschytz,
KLL} to the quantum mechanics.

The 0-brane action (\ref{action}) has a manifest $G_2$ global
symmetry, but the expectation value (\ref{vev}) breaks this to
$SU(3)$, so we begin by rewriting the action in form with manifest
$SU(3)$ invariance.  Under $SU(3) \subset G_2$ the ${\bf 7}$ decomposes
into ${\bf 3} \oplus \bar{\bf 3} \oplus {\bf 1}$.  Thus in place of
the seven real scalar multiplets $\Phi_A$ we have a set of three
complex scalar multiplets $\Phi_a$ transforming in the ${\bf 3}$,
their adjoints $\Phi_\ba^\dagger$ in the $\bar{\bf 3}$, and a single
real scalar multiplet $\Phi_7$.  The 0-brane action reads
\bea
\label{SU3action}
S & = & {1 \over g^2_{YM}} \int dt d^2 \theta \, {\rm Tr} \biggl\lbrace
- D^\alpha \Phi_\ba^\dagger D_\alpha \Phi_a - {1 \over 2} D^\alpha \Phi_7 D_\alpha \Phi_7 \\
\nonumber
& & \qquad + {1 \over 3 \sqrt{2}} \epsilon_{abc} \Phi_a [\Phi_b,\Phi_c] - {1 \over 3 \sqrt{2}} \epsilon_{\ba\bb\bc}
\Phi_\ba^\dagger [\Phi_\bb^\dagger,\Phi_\bc^\dagger] + \Phi_7 [\Phi_a, \Phi_\ba^\dagger] \\
\nonumber
& & \qquad + \,\,\, \hbox{\rm terms involving the gauge connection} \biggr\rbrace~.
\eea
We expand about the background (\ref{vev}), setting
\bea
\nonumber
\Phi_a & = & \left(\begin{array}{cc}
\hat{\Phi}_a & \delta \Phi_a \\
\delta \tilde{\Phi}_a^\dagger & 0
\end{array}\right) \\
\label{fieldexpansion}
\Phi_\ba^\dagger & = & \left(\begin{array}{cc}
\hat{\Phi}_\ba^\dagger & \delta \tilde{\Phi}_\ba \\
\delta \Phi_\ba^\dagger & 0
\end{array}\right) \\
\nonumber
\Phi_7 & = & \left(\begin{array}{cc}
\hat{\Phi}_7 & \delta \Phi_7 \\
\delta \Phi_7^\dagger & R
\end{array}\right)
\eea
with a similar expansion for the gauge connection.  The hatted fields
are $N \times N$ matrices describing the black hole background, while
the off-diagonal fields describe W-bosons in the fundamental of
$U(N)$, and $R$ is the Higgs vev which parameterizes the position of
the probe.

Expanding the action in powers of the off-diagonal fields, we have
(with $\hat{\Phi}$ referring to all of the scalar multiplets as well as
the gauge connection)
\be
\label{actionexpansion}
S = S_{\rm background}(\hat{\Phi}) + S_{\rm quadratic}(\hat{\Phi},\delta \Phi) + \ldots\,.
\ee
The zeroth order terms describe the black hole background.  All first
order terms automatically vanish, given the matrix structure
(\ref{fieldexpansion}) and the fact that the action (\ref{SU3action})
involves a single overall trace.  We can stop with the quadratic terms
in (\ref{actionexpansion}), since the off-diagonal fields transform in
the fundamental of $U(N)$, which means the higher-order terms make
contributions that are suppressed by $1/N$ in the large-$N$ limit.

To get a tractable description of the black hole background, we use
our mean-field approximation scheme \cite{KLL}.  In this approximation
one constructs a trial action $S_0(\hat{\Phi})$, which can be thought
of as a variational approximation to the full action $S_{\rm
background}(\hat{\Phi})$.  We took $S_0$ to essentially be the most
general quadratic action that one can write in terms of the
fundamental background fields.  The trial action is therefore
characterized by a set of 2-point functions, which were obtained by
solving a set of truncated Schwinger-Dyson equations.

To make a mean-field approximation for the background, we replace
$S_{\rm background} \rightarrow S_0$ in (\ref{actionexpansion}).  This
gives us an effective description of the probe, in terms of the action
\be
\label{action2}
S = S_0(\hat{\Phi}) + S_{\rm quadratic}(\hat{\Phi},\delta \Phi)\,.
\ee
In principle this action can be solved by standard large-$N$ techniques, since $S_0$ is a known
quadratic action and the fields $\delta \Phi$ are in the fundamental
of $U(N)$.  However, for simplicity, we wish to make a further
approximation: we will only keep fields which transform in the ${\bf
3}$ or $\bar{\bf 3}$ of $SU(3)$.  We will say more
about the validity of this truncation in section 5.  Given these
approximations, the action of interest is explicitly given by
\bea
\label{action3}
S & = & S_0(\hat{\Phi}_a,\hat{\Phi}^\dagger_\ba) + {1 \over g^2_{YM}} \int dt d^2 \theta \, \biggl\lbrace
- D^\alpha \delta \Phi_\ba^\dagger D_\alpha \delta \Phi_a - R \delta \Phi_\ba^\dagger \delta \Phi_a \\
\nonumber
& & \quad - D^\alpha \delta \tilde\Phi_a^\dagger D_\alpha \delta \tilde{\Phi}_\ba
+ R \delta \tilde{\Phi}_a^\dagger \delta \tilde{\Phi}_\ba
+ \sqrt{2} \epsilon_{abc} \delta \tilde{\Phi}_a^\dagger \hat{\Phi}_b \delta\Phi_c
- \sqrt{2} \epsilon_{\ba\bb\bc} \delta \Phi_\ba^\dagger \hat{\Phi}_\bb^\dagger \delta \tilde{\Phi}_\bc \biggr\rbrace\,.
\eea

We are interested in studying this theory at finite temperature.  To
do this we use an imaginary-time formalism.  We adopt the component
expansions discussed in appendix A
\beas
\delta \Phi_a & = & \delta \phi_a + i \delta \psi_{\alpha a} \theta_\alpha + \delta f_a \theta^2 \\
\delta \Phi_\ba^\dagger & = & \delta \phi_\ba^\dagger + i \delta \psi_{\alpha \ba}^\dagger \theta_\alpha + \delta f_\ba^\dagger \theta^2
\eeas
(same for tilded fields).  We continue to Euclidean space by setting
\[
S_E = - i S_M \qquad \tau = i t_M \qquad f_E = -i f_M\,.
\]
Note that the auxiliary fields must be Wick rotated to obtain a
Euclidean action that is bounded below.  (The rotation is somewhat
subtle; we need both $\delta f_{aE} = -i \delta f_{aM}$ and $\delta f_{\ba
E}^\dagger = -i \delta f_{\ba M}^\dagger$).  We then compactify
Euclidean time on a circle of circumference $\beta$, and expand the
fields in Matsubara modes.  For example we write
\beas
\delta \phi_a(\tau) & = & {1 \over \sqrt{\beta}} \sum_{l \in {\mathbb Z}} \delta \phi_{al} e^{i 2 \pi l \tau / \beta} \\
\delta \psi_{\alpha a}(\tau) & = & {1 \over \sqrt{\beta}} \sum_{r \in {\mathbb Z} + 1/2} \delta \psi_{\alpha a r} e^{i 2 \pi r \tau / \beta}\,.
\eeas

The model (\ref{action3}) can be solved by standard large-$N$
methods.  The leading contribution to the free energy is ${\cal
O}(N^2)$, and comes from the black hole background.  The leading
contribution to the free energy of the probe is ${\cal O}(N)$, and
comes from a single loop of W-bosons.  Since the action
(\ref{action3}) only has 3-point couplings involving two W-bosons and
one background field, the W propagator is given exactly by summing rainbow
diagrams, just as in the 't Hooft model of two-dimensional
chromodynamics \cite{tHooft}.  The mean-field methods used in
\cite{KabatLifschytz, KLL} therefore provide an exact description of
the probe, and we shall state the solution to the model using the
language of mean-field theory.

At leading order in $1/N$ the properties of the probe are completely
characterized by the W propagators.  We denote these propagators
\bea
\nonumber
\langle\delta \phi^\dagger_{\ba l I} \delta \phi_{b m J}\rangle_0 & = & \Delta_l^2 \delta_{\ba b} \delta_{lm} \delta_{IJ} \\
\label{propagators}
\langle\delta f^\dagger_{\ba l I} \delta \phi_{b m J}\rangle_0 & = & \langle\delta \phi^\dagger_{\ba l I} \delta f_{b m J}\rangle_0
= i \gamma_l \delta_{\ba b} \delta_{lm} \delta_{IJ} \\
\nonumber
\langle\delta f^\dagger_{\ba l I} \delta f_{b m J}\rangle_0 & = & \epsilon_l^2 \delta_{\ba b} \delta_{lm} \delta_{IJ} \\
\nonumber
\langle\delta \psi^\dagger_{\alpha \ba r I} \delta \psi_{\beta b s J}\rangle_0  & = & i g_r \delta_{\alpha \beta} \delta_{\ba b} \delta_{rs} \delta_{IJ}
+ i h_r \epsilon_{\alpha\beta} \delta_{\ba b} \delta_{rs} \delta_{IJ}
\eea
(same for tilded fields and propagators) where $I,J$ are indices in
the fundamental of $U(N)$.  These propagators are to be determined by
solving a set of one-loop gap equations, which we give below.  The
background action $S_0(\hat{\Phi})$ is also characterized by a set of
2-point functions, which we denote\footnote{The propagators
$\hat{\Delta}^2_l$ and $\hat{\epsilon}^2_l$ are the same as in
\cite{KLL}, but $\hat{g}_r$ is $i$ times the fermion propagator of
\cite{KLL}.}
\beas
\langle\hat{\phi}^\dagger_{\ba l IJ} \hat{\phi}_{b m KL}\rangle_0 & = & \hat{\Delta}_l^2 \delta_{\ba b} \delta_{lm} \delta_{IL} \delta_{JK} \\
\langle\hat{f}^\dagger_{\ba l IJ} \hat{f}_{b m KL}\rangle_0 & = & \hat{\epsilon}_l^2 \delta_{\ba b} \delta_{lm} \delta_{IL} \delta_{JK} \\
\langle\hat{\psi}^\dagger_{\alpha \ba r IJ} \hat{\psi}_{\beta b s KL}\rangle_0 & = &
i \hat{g}_r \delta_{\alpha \beta} \delta_{\ba b} \delta_{rs} \delta_{IL} \delta_{JK}\,.
\eeas
The action (\ref{action3}) has a ${\mathbb Z}_2$ symmetry which
exchanges $\delta \Phi$ and $\delta \tilde{\Phi}$, and takes $R
\rightarrow - R$, $\hat{\Phi} \rightarrow - \hat{\Phi}$.  This
symmetry implies that
\be
\Delta_l^2 = \tilde{\Delta}_l^2 \qquad \gamma_l = - \tilde{\gamma}_l \qquad \epsilon_l^2 = \tilde{\epsilon}_l^2
\qquad g_r = \tilde{g}_r \qquad h_r = - \tilde{h}_r\,.
\ee
From now on we will use this relation to eliminate the tilded propagators.

The propagators appearing in (\ref{propagators}) are to be determined
by solving a set of one-loop gap equations.  The gap equations can be
obtained by demanding that the two-loop 2PI effective action discussed
in \cite{KabatLifschytz, KLL} is stationary with respect to
infinitesimal variations of the propagators.  The free energy of the
probe can then be obtained by evaluating the effective action at the
critical point.  In the present case, after a rescaling discussed
below, the effective action $\beta F$ is given by
\bea
\label{betaF}
& & \beta F = - 6 \sum_l \log \left( \Delta_l^2 \epsilon_l^2 + (\gamma_l)^2\right) + 6 \sum_r \log \left( (g_r)^2 + (h_r)^2 \right) \\
\nonumber
& & + 6 \sum_l \left( ({2 \pi l \over \beta})^2 \Delta_l^2 - 2 R \gamma_l + \epsilon_l^2 - 2 \right)
- 12 \sum_r \left( {2 \pi r \over \beta} g_r - R h_r - 1 \right) \\
\nonumber
& & + {24 \over \beta} \sum_{l + m + n = 0} \Delta_l^2 \hat{\Delta}_m^2 \epsilon_n^2
+ {12 \over \beta} \sum_{l + m + n = 0} \Delta_l^2 \hat{\epsilon}_m^2 \Delta_n^2
+ {24 \over \beta} \sum_{l + m + n = 0} \gamma_l \hat{\Delta}_m^2 \gamma_n \\
\nonumber
& & + {48 \over \beta} \sum_{l + r + s = 0} g_r \hat{g}_s \Delta_l^2
+ {24 \over \beta} \sum_{l + r + s = 0} g_r \hat{\Delta}_l^2 g_s
- {24 \over \beta} \sum_{l + r + s = 0} h_r \hat{\Delta}_l^2 h_s~.
\eea
The gap equations and effective action are discussed in more detail in
appendix B.

Let us briefly note an important feature of our solution.  In
evaluating the effective action, we only kept planar diagrams.  This
makes 't Hooft large-$N$ counting automatic, so the free energy of the
probe is guaranteed to be ${\cal O}(N)$.  Moreover, the Yang-Mills
coupling constant can only appear in the combination $g^2_{YM} N$.  In
(\ref{betaF}), and in the rest of this paper, we suppress the overall
factor of $N$ in the probe free energy.  Also, by rescaling all
dimensionful quantities as in appendix B, we effectively adopt units
which set $g^2_{YM} N = 1$.

To solve the gap equations we used the numerical methods described in
\cite{KLL}, which we will not review here.  The basic idea is to start
by solving the gap equations at large $R$, where the theory is weakly
coupled, and then use the Newton-Raphson method to solve the gap equations at
a sequence of successively smaller radii.

%%%%%%%%%%%%%%%%%%%%%%%%%%%%%%%%%%%%%%%%%%%%%%%%%%%%%%%%%%%%%%%%%%%%%%%%%%%%%%
\section{Spectral analysis}
%%%%%%%%%%%%%%%%%%%%%%%%%%%%%%%%%%%%%%%%%%%%%%%%%%%%%%%%%%%%%%%%%%%%%%%%%%%%%%

At leading order in $1/N$ the properties of the probe are completely
characterized by the W propagators.  In the previous section we
described how these are computed numerically, as functions of
imaginary time.  There are a number of interesting questions that are
difficult to address, however, simply given the imaginary time
propagators.  For example, we would like to determine the masses of
the W-bosons, to see if supergravity predictions for the behavior of
strings that stretch between the probe and horizon are borne out, and
to fix the relation between the supergravity radial coordinate $U$ and
the Higgs expectation value $R$.  We would also like to determine the
entropy of the probe in a fixed black hole background.  These things
are not easily done in imaginary time.

To address these questions, it is useful to introduce a spectral
representation for the W propagators.  By inserting complete sets of
states one can derive the following analog of the Lehmann-K\"allen
spectral representation for a scalar propagator
\be
\label{SpectralRep}
\langle \phi(\tau) \phi(0) \rangle_\beta = \int_0^\infty d\omega \rho(\omega)
{\cosh \omega(\tau - \beta/2) \over 2 \omega \sinh (\beta \omega / 2)}
\qquad \quad \hbox{\rm $0 \leq \tau \leq \beta$}~.
\ee
Here the spectral density is given by
\[
\rho(\omega) = {1 \over Z} \sum_m e^{-\beta E_m} \sum_{n > m} \vert \langle n \vert \phi \vert m\rangle \vert^2\,\,
2 \omega \left(1 - e^{-\beta\omega}\right) \delta(\omega - E_n + E_m)\,.
\]
We will apply this spectral decomposition to the fields $\delta
\phi_a$, writing their momentum-space propagator as
\be
\label{SpectralRep2}
\Delta_l^2 = \int_0^\infty d\omega \rho(\omega) {1 \over (2 \pi l / \beta)^2 + \omega^2}\,.
\ee
We can regard $\rho(\omega) d\omega$ as the number of single-string
microstates having an energy between $\omega$ and $\omega + d
\omega$.\footnote{In Minkowski space the propagator has a branch cut
along the support of $\rho$.  Ordinarily this would reflect
multi-particle intermediate states.  In the case at hand the branch
cut arises because there are $N$ different W-bosons which can be
created by the operator $\delta \phi_a$, and in the large-$N$ limit these
W-bosons have a continuous distribution of masses.}  By expanding
the propagator at large momentum as discussed in appendix B,
note that the spectral density should satisfy
\be
\label{normalization}
\int d\omega \rho(\omega) = 1 \qquad
\int d\omega \rho(\omega) \omega^2 = m_\Delta^2
\ee
where $m_\Delta^2$ is the asymptotic mass (\ref{AsymptoticMass}).

An interesting problem, motivated by our discussion of supergravity in
section 2, is to determine the energies of the lightest and heaviest
string states $\omega_{\rm min}$ and $\omega_{\rm max}$.  It is easy
to put bounds on these quantities, as follows.  From
(\ref{normalization}) we have
\[
m_\Delta^2 = \int d\omega \rho(\omega) \omega^2 \leq \omega_{\rm max}^2\,.
\]
Also one can use (\ref{SpectralRep}) to show that
\be
\label{OmegaMinBound}
\cosh(\beta\omega_{\rm min}/2) \leq {\langle \phi(0) \phi(0) \rangle_\beta \over
\langle \phi(\beta/2) \phi(0) \rangle_\beta}\,.
\ee
Thus we have an upper bound on $\omega_{\rm min}$, and a lower bound
on $\omega_{\rm max}$.  These bounds are shown in Fig.~1.  Note that
as the temperature decreases, $\omega_{\rm min}$ develops a rather
sharp plateau at small radius.  We take this to indicate that the
probe has started to come to equilibrium with the black hole, in the
range of Higgs vevs corresponding to the plateau.

\begin{figure}
\epsfig{file=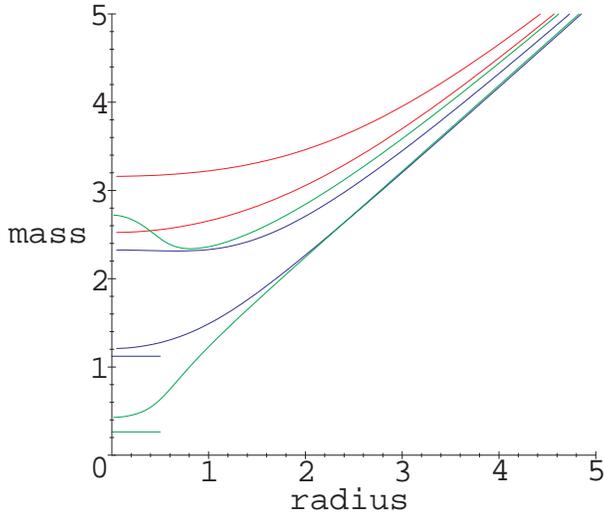}
\caption{Bounds on $\omega_{\rm min}$ and $\omega_{\rm max}$ vs.~$R$.
Red: $\beta = 0.1$ (perturbative regime), blue: $\beta = 0.9$, green: $\beta
= 3.8$.  The short horizontal lines mark the corresponding
temperatures.}
\end{figure}

Let us point out a few features of these results, assuming that the
actual values of $\omega_{\rm min}$ and $\omega_{\rm max}$ are close
to saturating the bounds we have derived.  At large radius the masses
are roughly given by the classical formula $m_W = R$.  But at small
radius and low temperature, we see clear evidence both for very light
states, with a mass of order the temperature, and for heavy states,
with a mass of order the 't Hooft scale $(g^2_{YM} N)^{1/3}$.  The
light states are expected, based on our discussion of supergravity in
section 2.  The heavy states are the degrees of freedom which must be
integrated out, as in section 3.1, in order to recover a local
supergravity description.

We now turn to the problem of directly determining the spectral
density.  In principle $\rho(\omega)$ is uniquely determined, given
the momentum-space propagators evaluated at an infinite number of
Matsubara frequencies and some assumptions about the behavior of the
propagators at infinity.  In practice, however, it is difficult to
determine $\rho(\omega)$ by inverting (\ref{SpectralRep2}).  The
integration over frequency smooths out features present in
$\rho(\omega)$. Consequently the inversion process has the opposite
effect, and suffers from numerical instability.

\begin{figure}
\epsfig{file=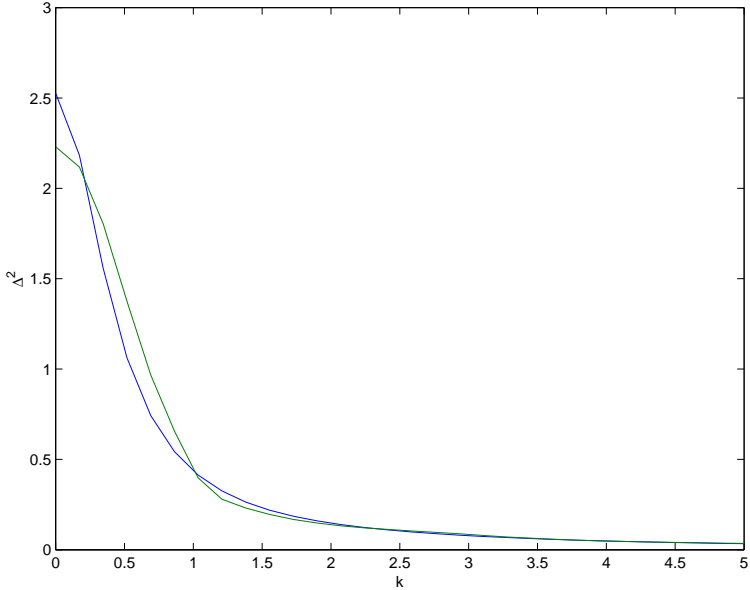,width=10cm}
\caption{The propagator $\Delta^2(k)$, obtained by solving
(\ref{ContinueGap}) at $\beta = 3.0$ and $R = 0.25$.  The blue curve
is the continued propagator.  The green curve is reconstructed from the spectral
density.}
\end{figure}

To obtain results for the spectral density we used the following
prescription.  First we continued the gap equations (\ref{gap1}) --
(\ref{gap5}) to general Euclidean momenta, by writing, for example
\bea
\label{ContinueGap}
&&{\epsilon^2(k) \over \Delta^2(k) \epsilon^2(k) + \bigl(\gamma(k)\bigr)^2} = k^2
+ {4 \over \beta} \sum_l \hat{\Delta}_l^2 \epsilon^2\bigl(-k - {2 \pi l \over \beta}\bigr) \\
\nonumber
&& \qquad\qquad\qquad\qquad + {4 \over \beta} \sum_l \hat{\epsilon}_l^2 \Delta^2\bigl(-k - {2 \pi l \over \beta}\bigr)
+ {8 \over \beta} \sum_r \hat{g}_r g\bigl(-k-{2 \pi r \over \beta}\bigr)\,.
\eea
Then we obtained solutions to these equations for $R \geq 0.25$ at
$\beta = 2.0$, $3.0$, $4.0$. Below this value of $R$, and for larger
values of $\beta$, it became difficult to obtain physical solutions to these
equations. Fig.~2 shows the results of this procedure. By obtaining
the propagators at frequencies intermediate to the Matsubara
frequencies, we are able to obtain $\rho(\omega)$ with a much finer
resolution than the spacing between Matsubara frequencies $2
\pi/\beta$.

\begin{figure}
\epsfig{file=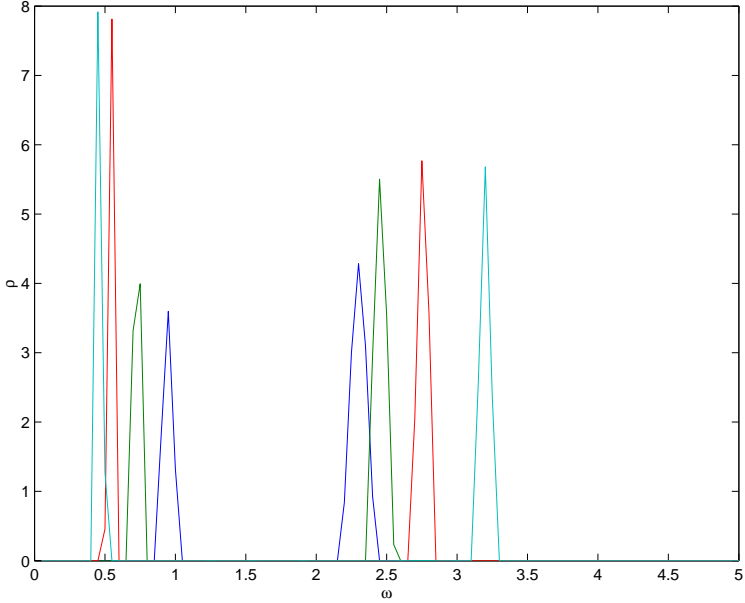,width=10cm}
\caption{The spectral density $\rho(\omega)$ at $\beta = 3.0$. Dark blue:
$R = 1.0$, green: $R = 0.75$, red: $R = 0.5$, light blue: $R = 0.25$.}
%\caption{The spectral density $\rho(\omega)$ at $\beta = 3.0$.  Red:
%$R = 1.0$, yellow: $R = 0.75$, green: $R = 0.5$, blue: $R = 0.25$.}
\end{figure}

\begin{figure}
\epsfig{file=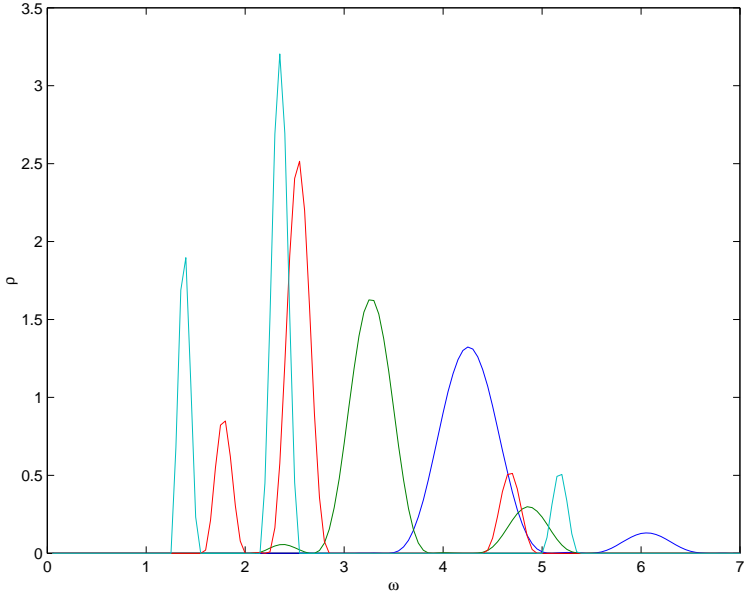,width=10cm}
\caption{The spectral density $\rho(\omega)$ at $\beta = 3.0$. Dark blue:
$R = 4.0$, green: $R = 3.0$, red: $R = 2.0$, light blue: $R = 1.5$.}
\end{figure}

Finally, to reconstruct the spectral densities, we used a constrained Tikhonov
regularization \cite{tikhonov}. The essential idea is to numerically
minimize a discretization of
\be
\label{chisq}
\sum_l \left\vert \Delta^2_l - \int_0^\infty d\omega ~\frac{ \rho(\omega)}{(2 \pi l
/\beta)^2 +\omega^2} \right\vert + \lambda^2 \int d\omega \left\vert \frac{d\rho}{d\omega} \right\vert\,.
\ee
The minimization is performed subject to the constraint that $\rho
\geq 0$. We also add terms corresponding to the constraints
(\ref{normalization}). The extra term dependent on the parameter
$\lambda$ makes the matrices that appear in the inversion process
well-conditioned. By choosing this extra term proportional to the norm
of the derivative of $\rho$ we enforce smoothness of the solution,
which helps to suppress numerical instabilities. We can actually choose this
term to be quite small $\lambda=10^{-4}$, so that the contribution of
the regulator to (\ref{chisq}) is negligible.  We should emphasize
that no starting ansatz is needed to perform the minimization, so no
prior knowledge about the final form of the solution is used as input,
other than the features already mentioned.  The spectral densities
that result are shown in Figs.~3 and 4.  

With this prescription, we believe we have obtained reliable results
for the spectral density.  The prescription seems to work best if the
radius is not too large; the small peaks seen at large $\omega$
($\omega \gtrsim 5$) in Fig.~4 may be numerical artifacts, since
perturbative quantum mechanics predicts a single peak at large $R$. However the
relative weight of these small high frequency peaks decreases as $R$
increases, consistent with perturbative expectations. More disturbing
is the fact that the dominant peak appears to get wider at large $R$,
in contrast to the perturbative result that there is a single
peak which becomes sharper as $R$ increases. This behavior appears to
be an artifact of the Tikhonov regularization. We have checked that
the large $R$ propagators can be well-fit by a single Lorentzian
spectral peak which becomes narrower as $R$ increases.
As a
check of our results, a comparison of the propagator reconstructed
from the spectral density and the original mean-field propagator is
shown in Fig.~2. The reason these do not agree more precisely is that
we have imposed a positivity constraint on the spectral density.  By
only partially summing Feynman diagrams, we have preserved unitarity
only approximately, so the spectral weight which would exactly
reproduce the mean field propagator need not be positive. The
reasonable agreement seen in Fig.~2 provides us with a good indicator
for how well the mean field approximation is working.

Let us make some comments on the spectral densities we have obtained,
neglecting the numerical artifacts which seem to be present for
$\omega \gtrsim 5$.  A striking feature of our results for the
spectral density is a bimodal distribution of W masses at sufficiently
small radii.  The results of \cite{KLL} indicate that the supergravity
regime should correspond to approximately $R\leq 2$.  At $R>2$ we find
a single dominant peak in the spectral density, consistent with what
is expected from the perturbative quantum mechanics.  However as $R$
decreases below the scale where supergravity becomes a good
approximation we find two peaks in the spectral density. This suggests
that once we enter this regime, two different types of fundamental
strings are contributing to the spectral weight. One set run between
the probe brane and the black hole horizon. These become light as $R$
decreases, and account for the entropy of the black hole in the limit
$R\to 0$. Another set of fundamental strings appear to run off to the
strongly curved asymptotic region of the supergravity background. As
can be seen from Fig.~3 the sum of the positions of the two peaks at a
given $R$ is roughly $3.5$, independent of $R$.  This is consistent
with the above interpretation.

\begin{figure}
\epsfig{file=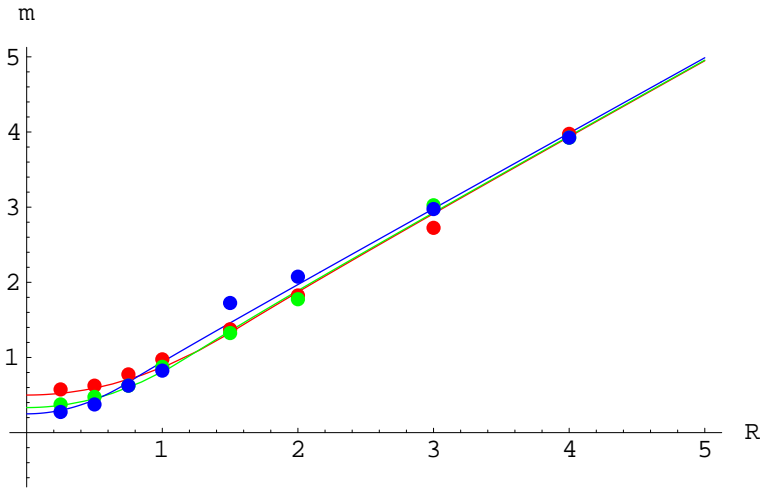}
\caption{The lightest $W$ mass as a function of $R$ at $\beta =
2.0$ (red), $\beta=3.0$ (green) and $\beta=4.0$ (blue). }
\end{figure}

A key feature of the spectral density is that a large number of light
$W$ boson states are present when $R$ is small.  We plot the lightest
$W$ mass as a function of $R$ in Fig.~5 (we define this as the
frequency $m_W$ which satisfies $\int_0^{m_W} d\omega \rho(\omega) =
0.1$).  At large $R$ we find $m_W \approx R$.  As $R$ decreases $m_W$
also decreases.  At some radius $m_W$ becomes of order the temperature
and thermalization starts to occur.  We identify this radius with the
stretched horizon of the black hole.  Note that the radius of the
stretched horizon decreases with temperature, as expected.

Our reconstruction of the spectral density shows that the energy of
the lightest W is approximately constant at small radius, for example
at $\beta = 3.0$ we find $m_W \approx 0.4$ for $R \lesssim 0.5$.  This
is compatible with the bound (\ref{OmegaMinBound}), which requires
$m_W < 0.51$ at $R = 0$.  This energy scale $m_W$ is indeed comparable
to the temperature $1/\beta = 0.33$.  Thus absorption of the probe by
the black hole can be understood as thermal restoration of $U(N+1)$
gauge symmetry.

%%%%%%%%%%%%%%%%%%%%%%%%%%%%%%%%%%%%%%%%%%%%%%%%%%%%%%%%%%%%%%%%%%%%%%%%%%%%%
\section{Probe potential: reconstructing the spacetime metric from
quantum mechanics}
%%%%%%%%%%%%%%%%%%%%%%%%%%%%%%%%%%%%%%%%%%%%%%%%%%%%%%%%%%%%%%%%%%%%%%%%%%%%%

We wish to compare the probe free energy to the supergravity potential.  To do this in a
meaningful way, we must first determine how the supergravity radial
coordinate $U$ is related to the Higgs expectation value $R$.  The
precise relationship can be obtained by comparing W masses, as
follows.

Consider the mass of the lightest W boson in the quantum mechanics.
At large $R$, where supersymmetry is restored and the quantum
mechanics is weakly coupled, we can determine W masses perturbatively;
a perturbative calculation gives $m_W = R + {\cal O}(1/R)$.  As $R$
becomes small, however, the spectral analysis of the previous section
indicates that the mass of the lightest W goes to a constant of order
the Hawking temperature $T$. To find a useful analytic form that
captures both these limits, we fit the lightest W mass to the
following ansatz.
\be
\label{mWansatz}
m_W = \left\lbrace
\begin{array}{ll}
T + a R^2 & \quad R < R_0 \\
\noalign{\vspace{2mm}}
R + b / R & \quad R > R_0
\end{array}
\right.
\ee
Here $R_0$ is an adjustable parameter, while the constants $a$ and $b$
are fixed by demanding continuity of $m_W(R)$ and its first derivative
at $R=R_0$.  Fitting the ansatz to the data points shown in Fig.~5
yields the interpolating curves also shown there, with $R_0=1.5$ at
$\beta=2.0$, $R_0=1.2$ at $\beta=3.0$, and $R_0=1.0$ at $\beta=4.0$.

On the other hand a supergravity computation of the W mass gives $m_W
= (U - U_0) / 2 \pi$ (\ref{mW}).  This is the mass of a string
which stretches from the probe to the horizon of the black hole.  It
seems appropriate to identify the mass of this string with the mass of
the lightest W boson in the quantum mechanics.  Thus we take the
relation between $U$ and $R$ to be
\be
\label{urrel}
{U - U_0 \over 2 \pi} = \left\lbrace
\begin{array}{ll}
T + a R^2 & \quad R < R_0 \\
\noalign{\vspace{2mm}}
R + b / R & \quad R > R_0
\end{array}
\right.
\ee
where $U_0$ is given by the supergravity relation (\ref{HawkingTemp}).
Of course this relation is only trustworthy in the region where
supergravity makes sense, {\em i.e.}~between the stretched horizon of
the black hole and the region of large curvature $U \sim (g^2_{YM}
N)^{1/3}$.

Let us make a few comments on this change of coordinates.
Our results for the potential will not require the ansatz (\ref{urrel}); we could
invert the relation between $m_W$ and $R$ given by the data in
Fig.~5, and express everything in terms of $m_W$.  However one may
wonder whether the ansatz (\ref{urrel}) captures the correct
functional relation between $U$ and $R$.  Another functional form has
been proposed in the literature \cite{ProbeAction,radial},
\be
R^{7/2} = {1 \over 2} \left( \biggl({U \over 2 \pi}\biggr)^{7/2} + \sqrt{\biggl({U \over 2 \pi}\biggr)^7 - \biggl({U_0 \over 2 \pi}\biggr)^7} \right)\,.
\ee
This predicts that the relation
\[
m_W = {U - U_0 \over 2 \pi} = {1 \over R} \left(R^7 + {1 \over 4} \biggl({U_0 \over 2 \pi}\biggr)^7\right)^{2/7} - {U_0 \over 2 \pi}
\]
should hold in the supergravity regime.  Suppose one makes an ${\cal
O}(T)$ correction to this formula, and takes
\be
\label{urrel2}
m_W = {1 \over R} \left(R^7 + {1 \over 4} \biggl({U_0 \over 2 \pi}\biggr)^7\right)^{2/7} - {U_0 \over 2 \pi} + T\,.
\ee
With $U_0$ regarded as an adjustable parameter, one can get a quite
good fit to the data shown in Fig.~5.  Within our numerical accuracy,
we cannot claim to distinguish between the two proposals (\ref{urrel})
and (\ref{urrel2}).

In Fig.~6 we show a plot of the probe free energy as a function of
$R$, obtained by evaluating (\ref{betaF}) at $\beta = 3.0$. The
continuous curve is the supergravity prediction for the effective
potential (\ref{SugraPotential}).  The $U$ coordinate is fixed using
the relation (\ref{urrel}) and the overall tension in the DBI action
(\ref{DBI}) is adjusted to fit the data.  Note that we do not expect the
overall coefficient to be accurately reproduced by the mean field
approximation -- in \cite{KLL}, where a complete mean-field
calculation of the free energy of the background was performed, there
was a $50\%$ discrepancy between the predicted coefficient and mean
field, although the scaling exponent with temperature was reproduced
to within a few percent.  In our probe mean-field calculations we have
not included the gauge multiplet or the 7${}^{th}$ scalar multiplet,
and these fields are expected to make a substantial contribution to
the overall coefficient.

Fig.~6 shows remarkable agreement for the shape of the potential. The
$U$ vs. $R$ relation (\ref{urrel}) is crucial in obtaining such good
agreement.  Because the supergravity effective potential depends in
such a detailed way on the form of the black hole metric, we see that
the mean field approximation in the quantum mechanics is accurately
reproducing the spacetime physics of the black hole geometry.

\begin{figure}
\epsfig{file=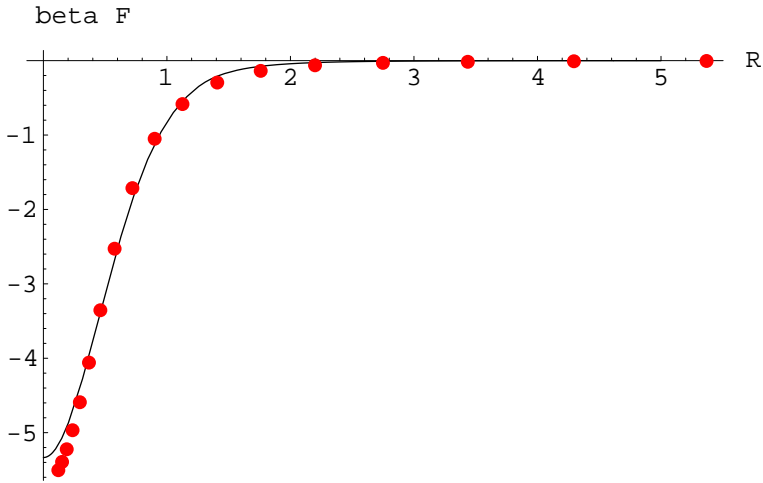}
\caption{The probe free energy $\beta F$ vs.~$R$, at $\beta =
  3.0$. The curve shows the Born-Infeld action (\ref{SugraPotential})
  with the overall tension fit to the data.}
\end{figure}

One caveat is worth mentioning.  The free energy of the probe
(\ref{betaF}) falls off like $1/R^3$ at large radius.  This behavior
is a consequence of the ${\cal N} = 2$ supersymmetry of the truncated
action (\ref{action3}).  Note that supergravity has a potential
(\ref{SugraPotential}) which falls off like $1/U^7$ far outside the
horizon (at $U \gg U_0$).  Thus with the truncation (\ref{action3}) we
could not hope to see agreement with the long-distance behavior of
supergravity.  Fortunately, in the temperature range we are studying,
the horizon radius $U_0$ is large enough that the shape of the
supergravity effective potential is dominated by the square root
singularity as $U\to U_0$.  It is not particularly sensitive to
whether $U^7$ appears or some other power of $U$ appears in the
potential.  This makes it possible for the truncated probe theory to
reproduce well the shape of the potential.

Finally, let us discuss the behavior at small radius.  The gauge
theory has the property that as $R \rightarrow 0$ the probe 0-brane
becomes indistinguishable from the $N$ 0-branes that make up the black
hole background.  This is clear from the form of the expectation value
(\ref{vev}).  This behavior is compatible with the properties of
supergravity discussed at the end of section 2: the free energy of a
probe that has come to equilibrium with the black hole can be obtained
from the free energy of the background, simply by shifting $N
\rightarrow N+1$.

\begin{figure}
\epsfig{file=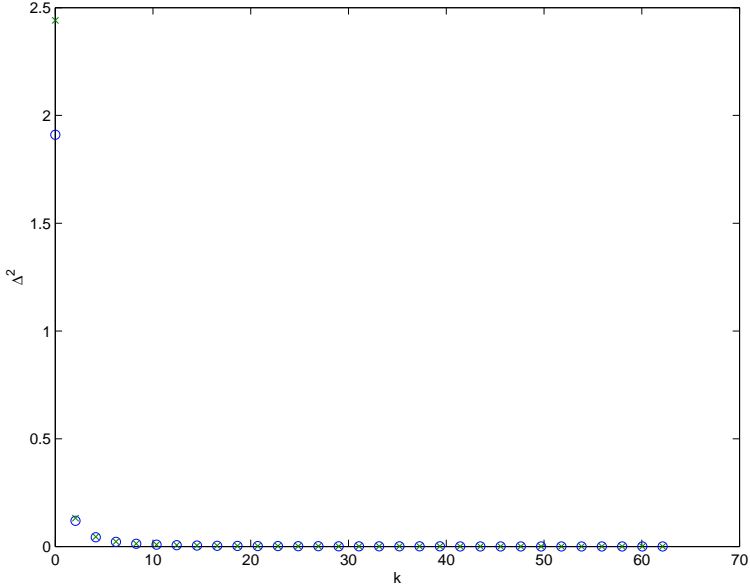,width=10cm}
\caption{Comparison of the scalar propagators at small $R$ in the full
mean field approximation (blue circles) and the truncated mean field
approximation (green x's), at $\beta = 3.0$.}
\end{figure}

This behavior is respected by our approximation, in the sense that as
$R \rightarrow 0$ the gap equations for the W propagators in the
theory (\ref{action2}) become identical to the one-loop gap equations
which we used to describe the black hole background in \cite{KLL}.
When we truncated the action, going from (\ref{action2}) to
(\ref{action3}), we slightly violated the property that the background
and W propagators agree at $R = 0$.  We can use this to test the
validity of the truncation.  A comparison of the scalar propagator at
$R=0.1$ in our truncation, and the corresponding propagator for the
background, is shown in Fig. 7. The zero frequency modes differ by
about $20\%$, however this discrepancy rapidly decreases for the
higher Matsubara modes: at $k=2\pi/\beta$ the difference is $9\%$, and
becomes less than $1\%$ for $k \geq 14\pi/\beta$.

The free energy is rather more sensitive to the truncation than the
propagators themselves. In the truncated probe mean field
approximation we find $\beta F_{\rm probe} = -5.7$ as $R \rightarrow
0$ at $\beta=3.0$, whereas a complete mean-field calculation would
give $\beta F_{\rm probe} = -2.0$ at $R=0$ (this follows from shifting
$N \rightarrow N+1$ in the results of \cite{KLL}).  These free
energies should be compared to the supergravity prediction
$\beta F_{\rm probe} = -0.80$ for a probe in equilibrium with the
black hole (\ref{shiftun}).  We see that the full mean field is off by a factor of 2.5
from the supergravity prediction, while the truncated probe mean field
is off by a factor of 7.  The dominant source of the discrepancy
between the two mean field results is the zero mode of the gauge
field. The fact that the shape of the curves in Fig.~6 match so well
suggests that the extra contribution due to the gauge zero mode leads only
to a renormalization of the probe mass, and does not alter the overall
shape of the potential. We plan to extend the probe mean field
approximation to include the gauge multiplet, as well as the
7${}^{th}$ scalar multiplet, in future work. This will be important
for seeing a more detailed matching of the probe mass, and also to
resolve finer features of the probe effective potential.

%%%%%%%%%%%%%%%%%%%%%%%%%%%%%%%%%%%%%%%%%%%%%%%%%%%%%%%%%%%%%%%%%%%%%%%%%%%%%%%%%%
\section{Probe Entropy}
%%%%%%%%%%%%%%%%%%%%%%%%%%%%%%%%%%%%%%%%%%%%%%%%%%%%%%%%%%%%%%%%%%%%%%%%%%%%%%%%%%

We now turn to computing the entropy of the probe as
a function of radius.  Our expectation is that outside the stretched
horizon, where supergravity is valid, the entropy of the probe should
be very small.  (In fact, according to classical supergravity, the
entropy of the probe is exactly zero.)  However as the probe
reaches the stretched horizon light W's become thermally
excited, and the probe acquires a non-zero entropy.  This should
provide a clear signal for the breakdown of supergravity at the
stretched horizon of the black hole.

We wish to compute the entropy of the probe while holding the
mass of the black hole (and the radius of the probe) fixed.  
In the imaginary time formalism, the temperature of the probe is tied
to the temperature of the black hole.  This means we cannot compute
the entropy using the canonical ensemble.  Rather we have to define the
probe entropy microcanonically.  This is easily done, given the
spectral representation (\ref{SpectralRep2}).  We merely have to
integrate the spectral density against the entropy of a harmonic
oscillator with frequency $\omega$ and temperature $1/\beta$
\be
\label{Sp}
S_{\rm probe} = N \int d\omega \rho(\omega) \left({\beta \omega \over 2 \tanh(\beta\omega/2)}
- \log\bigl(2 \sinh(\beta\omega/2)\bigr)\right)~.
\ee
The resulting entropy is shown in Fig.~8.  (This is the entropy of a
single W boson; to get the bosonic entropy of the full theory
(\ref{action3}) one should multiply by 6).  Of course there are also
fermionic strings, which make an (additive) contribution to the total
entropy, but the bulk of the entropy comes from the bosons.

We can see that the probe entropy is small at the edge of the
eigenvalue cloud.  For a spectral density of the sort we obtained in
section 4, (\ref{Sp}) predicts that the probe entropy is exponentially
small outside the stretched horizon, roughly $S_{\rm probe} \sim
e^{-\beta m_W} \sim e^{-\beta R}$.  It only begins to increase
dramatically at a considerably smaller radius, which we identify with
the stretched horizon of the black hole.  As can be seen in Fig.~8,
the stretched horizon moves to smaller values of $R$ as the
temperature decreases.  Also note that the entropy increases more
suddenly at lower temperatures.  Both these features are in accord
with the supergravity expectation (\ref{Us}).

\begin{figure}
\epsfig{file=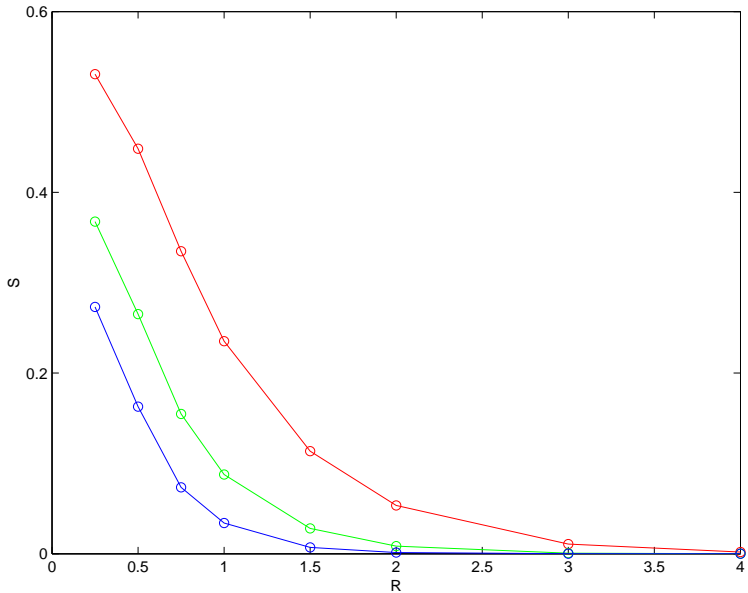,width=10cm}
\caption{Bosonic contribution to the probe entropy vs.~$R$ at $\beta =
2.0$ (red), $\beta=3.0$ (green) and $\beta=4.0$ (blue).  The edge of the eigenvalue
distribution (\ref{Wigner}) lies at about $R=1.8$.  
Numerical calculations were performed at the indicated points.}
\end{figure}

%%%%%%%%%%%%%%%%%%%%%%%%%%%%%%%%%%%%%%%%%%%%%%%%%%%%%%%%%%%%%%%%%%%%%%%%%%%%%%
\section{An entropy -- radius relation}
%%%%%%%%%%%%%%%%%%%%%%%%%%%%%%%%%%%%%%%%%%%%%%%%%%%%%%%%%%%%%%%%%%%%%%%%%%%%%%

As a final topic, we discuss an interesting relation between the
entropy of a black hole and the radius of the black hole horizon.  In
this section we do not use a 0-brane probe, although much of what
we will say is motivated by the results of the previous sections.

Within the context of our mean-field approximation \cite{KLL}, the
black hole is modeled as a collection of $N^2$ independent degrees of
freedom.  Let $\rho(\omega)$ be the corresponding spectral density for
the bosons.  Then the entropy of the black hole is given by a formula
analogous to (\ref{Sp}),
\be
\label{zeroth}
S_{\rm bh} \sim N^2 \int d\omega \rho(\omega) \left({\beta \omega \over 2 \tanh(\beta\omega/2)}
- \log\bigl(2 \sinh(\beta\omega/2)\bigr)\right)
\ee
where we have neglected a small contribution from the fermions.
The spectral density of the black hole background should be identical
to the spectral density of a probe at very small radius.  Having
analyzed the probe spectral density, we expect that $\rho(\omega)$ has
two peaks, one centered around $\omega \sim T$ and one centered around
much higher frequencies.  Moreover these peaks are very narrow.  So up
to a coefficient, the black hole entropy is just given by
the area of the first low-frequency peak
\be
\label{first}
S_{\rm bh} \sim N^2 \int_{\hbox{\rm first peak}} d\omega \rho(\omega)~.
\ee
Now let us obtain an expression for the radius of the black hole
horizon $R_h$.  We will define this, not in terms of a probe brane, but
rather as in section 5 of \cite{KLL}, in terms of the 2-point function
of the time-averaged scalar fields (\ref{SmearedOperator}).  In terms
of the spectral representation (\ref{SpectralRep}) we have
\be
R_h^2 \equiv \langle \bar{\phi}^2 \rangle = \int_0^{\Lambda} d\omega \rho(\omega) {1 \over 2 \omega \tanh(\beta \omega/2)}
\ee
where $\Lambda$ is a high-frequency cutoff, corresponding to a choice
of resolving time used to define the size of the black hole.
$\Lambda$ should be chosen to keep only the light modes, the modes which are
described by supergravity.  Since $\rho(\omega)$ has a bimodal
distribution, there is a natural set of frequencies to exclude; we take the
horizon size to be given by just integrating over the first low-frequency
peak.  As the peak is narrow and concentrated around $\omega \sim T$
we get (up to a numerical factor)
\be
\label{second}
R_h^2 \sim \beta \int_{\hbox{\rm first peak}} d\omega \rho(\omega)\,.
\end{equation}
Combining (\ref{first}) and (\ref{second}) we get a non-trivial relationship between the horizon radius
and entropy of the black hole.  Restoring units
\be
\label{third}
S_{\rm bh} \sim {N^2 T R_h^2 \over g^2_{YM} N}\,.
\ee
Note that we have obtained this relationship strictly within the gauge theory.

Remarkably, however, a supergravity relationship of this form is valid
for all black holes that arise as the near-horizon geometry of black
p-branes.  The supergravity relationship is
\be
\label{fourth}
S_{\rm bh} \sim {N^2 T U_0^2 \over g^2_{YM} N}
\ee
where the constant of proportionality depends on the dimension of the
brane.  This relationship was noticed in appendix C.2 of \cite{KabatLifschytz}.

There are a few points worth mentioning about this derivation.
\begin{enumerate}
\item
The gauge theory measures the horizon radius $R_h$ in terms of a Higgs
field, while supergravity measures the horizon radius $U_0$ in terms
of the radial coordinate $U$.  The two coordinates are not the same.
However in the range of temperatures we have studied, the two
coordinates do not differ significantly.  Also the considerations of
\cite{ProbeAction,radial} suggest that $R_h$ is always proportional to
$U_0$.
\item
The expression (\ref{zeroth}) is only the leading expression for the
entropy in the mean-field approximation.  There is an infinite series
of perturbative corrections to the leading mean-field result.  As
suggested in \cite{KabatLifschytz}, these higher corrections may only
change numerical coefficients, which we have anyways ignored.
\end{enumerate}
We take the agreement between (\ref{third}) and (\ref{fourth}) to mean
that the assumptions which went in to deriving (\ref{third}) are
qualitatively correct.  In particular, this supports our claim that
the spectral distribution has a clear separation of light and heavy degrees of freedom, and has a narrow peak at
frequencies $\omega \sim T$.

%%%%%%%%%%%%%%%%%%%%%%%%%%%%%%%%%%%%%%%%%%%%%%%%%%%%%%%%%%%%%%%%%%%%%%%%%%%%%%
\section{Conclusions}
%%%%%%%%%%%%%%%%%%%%%%%%%%%%%%%%%%%%%%%%%%%%%%%%%%%%%%%%%%%%%%%%%%%%%%%%%%%%%%

In this paper we studied a 0-brane probe of a ten-dimensional
non-extremal black hole, directly in terms of the dual
strongly-coupled quantum mechanics.  We described the black hole
background using the mean-field methods of \cite{KLL}.  Following
Susskind \cite{Susskind}, we showed that a localized probe could be
described in the quantum mechanics, once a suitable resolving time has
been introduced.  We studied the spectral representation of the W
propagators, and found that light states are present in the quantum
mechanics whenever the probe is inside the stretched horizon.  These
light states provide a mechanism for the black hole to absorb
infalling matter, along the lines suggested in \cite{causality}.  It
would be quite interesting if these light states could be related to
the light fractionated monopoles which Mathur proposed as an
absorption mechanism for certain black holes \cite{Mathur}.

Given the W propagators, it was straightforward to compute the free
energy of the probe.  We showed that outside the stretched horizon the
probe potential was in accord with supergravity expectations.  However
the probe acquires a non-zero entropy once it enters the stretched
horizon, as the light W states become thermally excited.  This
provides a clear signal that supergravity breaks down at the stretched
horizon of a black hole, at least according to a Schwarzschild
observer.

There are several interesting directions in which one could extend the
results of this paper.  For example, for simplicity we studied a
truncated model for the probe (\ref{action3}), in which several fields
were suppressed.  But one can solve the full model (\ref{action2}),
using essentially the same techniques.  This should lead to improved
results, in particular for the probe effective potential.  Another
interesting problem would be to study a probe with non-zero velocity.
One could then compute, not only the probe potential, but also the
probe kinetic terms.  One could then hope to identify the non-trivial
causal structure of the black hole metric (\ref{metric}), as reflected
in the probe effective action (\ref{probeDBI}).

\bigskip
\goodbreak
\centerline{\bf Acknowledgements}
\noindent
We are grateful to Mike Douglas, Emil Martinec and Lenny Susskind for valuable
discussions.  NI and DK are supported by the DOE under contract
DE-FG02-92ER40699.  GL would like to thank Columbia University
and Tel-Aviv University for hospitality.  GL is supported in
part by US--Israel binational science foundation grant 2000359.  The
research of DL is supported in part by DOE grant DE-FE0291ER40688 Task A.

\appendix
%%%%%%%%%%%%%%%%%%%%%%%%%%%%%%%%%%%%%%%%%%%%%%%%%%%%%%%%%%%%%%%%%%%%%%%%%%%%%%
\section{${\cal N} = 2$ superspace}
%%%%%%%%%%%%%%%%%%%%%%%%%%%%%%%%%%%%%%%%%%%%%%%%%%%%%%%%%%%%%%%%%%%%%%%%%%%%%%

With ${\cal N} = 2$ supersymmetry we have an $SO(2)$ R-symmetry, with
spinor indices $\alpha,\beta = 1,2$ and vector indices $i,j = 1,2$.
The $SO(2)_R$ Dirac matrices $\gamma^i_{\alpha\beta}$ are real,
symmetric, and traceless.  Given two spinors $\psi$ and $\chi$, there
are two invariants one can make, which we denote by
\[
\psi_\alpha \chi_\alpha \qquad {\rm and} \qquad \psi^\alpha
\chi_\alpha \equiv {i \over 2} \epsilon_{\alpha\beta}
\psi_\alpha \chi_\beta\,.
\]
${\cal N} = 2$ superspace has coordinates $(t,\theta_\alpha)$, where
$\theta_\alpha$ is a collection of real Grassmann variables that
transform as a spinor of $SO(2)_R$.  The simplest representation of
supersymmetry is a real scalar superfield (complex conjugation reverses
the order of Grassmann variables)
\[
\Phi = \phi + i \psi_\alpha \theta_\alpha + f \theta^2\,.
\]
It contains a physical real boson $\phi$ and a physical real fermion
$\psi_\alpha$, along with a real auxiliary field $f$.  To describe
gauge theory we introduce a real spinor connection on superspace
$\Gamma_{\alpha}$, with component expansion
\[
\Gamma_\alpha = \chi_\alpha + A_0 \theta_\alpha + X^i
\gamma^i_{\alpha\beta} \theta_\beta + d \epsilon_{\alpha\beta}
\theta_\beta + 2 \epsilon_{\alpha\beta} \lambda_\beta \theta^2 \,.
\]
The fields $X^i$ are physical scalars, while $\lambda_\alpha$ are
their superpartners, $d$ is an auxiliary boson, $\chi_\alpha$ are
auxiliary fermions, and $A_0$ is the 0+1 dimensional gauge field.

To write a Lagrangian we introduce a supercovariant derivative
\be
D_\alpha  =  {\partial \over \partial \theta_\alpha} - i \theta_\alpha
{\partial \over \partial t}
\ee
and its gauge-covariant extension
\be
\nabla_\alpha = D_\alpha + \Gamma_\alpha\,.
\ee
The action for $N$ 0-branes is built from a collection of seven adjoint scalar
multiplets $\Phi_A$ that transform in the ${\bf 7}$ of a $G_2 \subset
SO(9)$ global symmetry, coupled to a $U(N)$ gauge multiplet
$\Gamma_\alpha$.  The action reads
\be
\label{SYMaction}
S = {1 \over g^2_{YM}} \int dt d^2\theta \, {\rm Tr} \biggl(
- {1 \over 4} \nabla^\alpha {\cal F}_i \nabla_\alpha {\cal F}_i
- \frac{1}{2} \nabla^\alpha \Phi_A \nabla_\alpha \Phi_A
- \frac{i}{3} f_{ABC} \Phi_A [\Phi_B, \Phi_C] \biggr)\,.
\ee
Here ${\cal F}_i = {1 \over 4} \gamma^i_{\alpha\beta} \lbrace
\nabla_\alpha, \nabla_\beta \rbrace$ is the field strength constructed
from $\Gamma_\alpha$, and $f_{ABC}$ is a totally
antisymmetric $G_{2}$-invariant tensor, normalized to
satisfy
\be
f_{ABC} f_{ABD} = {3 \over 2} \delta_{CD}\,.
\ee

%%%%%%%%%%%%%%%%%%%%%%%%%%%%%%%%%%%%%%%%%%%%%%%%%%%%%%%%%%%%%%%%%%%%%%%%%%%%%%%%
\section{Effective action and gap equations}
%%%%%%%%%%%%%%%%%%%%%%%%%%%%%%%%%%%%%%%%%%%%%%%%%%%%%%%%%%%%%%%%%%%%%%%%%%%%%%%%

The propagators (\ref{propagators}) correspond to a Gaussian trial
action for the off-diagonal fields (with a similar action for tilded fields)
\bea
\label{trial}
S_0(\delta \Phi, \delta \Phi^\dagger) & = & \sum_{a,l} \left(
\begin{array}{cc}
\delta \phi_{\ba l}^\dagger & \delta f_{\ba l}^\dagger
\end{array}
\right)
\left(
\begin{array}{cc}
\Delta_l^2 & i \gamma_l \\ i \gamma_l & \epsilon_l^2
\end{array}
\right)^{-1}
\left(
\begin{array}{c}
\delta \phi_{a l} \\ \delta f_{a l}
\end{array}
\right) \\
\nonumber
& & + i \sum_{a,r} \left(
\begin{array}{cc}
\delta\psi_{1 \ba r}^\dagger & \delta\psi_{2 \ba r}^\dagger
\end{array}
\right)
\left(
\begin{array}{cc}
g_r & - h_r \\ h_r & g_r
\end{array}
\right)^{-1}
\left(
\begin{array}{c}
\delta \psi_{1 a r} \\ \delta \psi_{2 a r}
\end{array}
\right)\,.
\eea
The 2-loop 2PI effective action discussed in \cite{KabatLifschytz,
KLL} is defined by\footnote{This quantity was denoted $I_{\rm eff}$ in
\cite{KLL}.}
\[
\beta F  = \beta F_0 + \langle S - S_0 \rangle_0 - {1 \over 2} \langle (S_{III}^2) \rangle_{\C,0}
\]
where $\beta F_0$ is the free energy of the trial action
(\ref{trial}), $S - S_0$ is the difference between the full action
(\ref{action3}) and the trial action (\ref{trial}), and $S_{III}$
refers to cubic terms in the full action.  A subscript $\bigC,0$ denotes
a connected expectation value computed using the trial action.  In the present case,
the effective action is given by
\bea
\label{betaF2}
& & \beta F = - 6 N \sum_l \log \left( \Delta_l^2 \epsilon_l^2 + (\gamma_l)^2\right) + 6 N \sum_r \log \left( (g_r)^2 + (h_r)^2 \right) \\
\nonumber
& & + {6 N \over g^2_{YM}} \sum_l \left( ({2 \pi l \over \beta})^2 \Delta_l^2 - 2 R \gamma_l + \epsilon_l^2 - 2 g^2_{YM}\right)
- {12 N \over g^2_{YM}} \sum_r \left( {2 \pi r \over \beta} g_r - R h_r - g^2_{YM} \right) \\
\nonumber
& & + {24 N^2 \over g^4_{YM} \beta} \sum_{l + m + n = 0}
\Delta_l^2 \hat{\Delta}_m^2 \epsilon_n^2 + {12 N^2 \over g^4_{YM} \beta} \sum_{l + m + n = 0} \Delta_l^2 \hat{\epsilon}_m^2 \Delta_n^2
+ {24 N^2 \over g^4_{YM} \beta} \sum_{l + m + n = 0} \gamma_l \hat{\Delta}_m^2 \gamma_n \\
\nonumber
& & + {48 N^2 \over g^4_{YM} \beta} \sum_{l + r + s = 0} g_r \hat{g}_s \Delta_l^2 + {24 N^2 \over g^4_{YM} \beta} \sum_{l + r + s = 0}
g_r \hat{\Delta}_l^2 g_s - {24 N^2 \over g^4_{YM} \beta} \sum_{l + r + s = 0} h_r \hat{\Delta}_l^2 h_s~.
\eea
The effective action respects 't Hooft large-$N$ scaling, so all
factors of $g^2_{YM} N$ can be eliminated from (\ref{betaF2}) by
appropriate rescalings of dimensionful quantities.  For example,
one sets
\[
\beta = \beta' / (g^2_{YM} N)^{1/3} \quad R = R' (g^2_{YM} N)^{1/3} \quad \Delta_l^2 = (g^2_{YM} N)^{1/3} \Delta'_l{}^2 / N\,.
\]
The rescaled effective action, with the overall factor of $N$
suppressed, is given in (\ref{betaF}).  Requiring that the rescaled effective
action is stationary with respect to variation of the propagators
gives rise to the following set of gap equations
\bea
\nonumber
&&{\epsilon_l^2 \over \Delta_l^2 \epsilon_l^2 + (\gamma_l)^2} = (\kl)^2 + {4 \over \beta} \sum_{m+n = -l} \hat{\Delta}_m^2 \epsilon_n^2
+ {4 \over \beta} \sum_{m+n = -l} \hat{\epsilon}_m^2 \Delta_n^2 + {8 \over \beta} \sum_{r+s = -l} \hat{g}_r g_s \\
\label{gap1}
& & \\
\label{gap2}
&&{\Delta_l^2 \over \Delta_l^2 \epsilon_l^2 + (\gamma_l)^2} = 1 + {4 \over \beta} \sum_{m+n = -l} \hat{\Delta}_m^2 \Delta_n^2 \\
\label{gap3}
&&{\gamma_l \over \Delta_l^2 \epsilon_l^2 + (\gamma_l)^2} = - R + {4 \over \beta} \sum_{m+n = -l} \hat{\Delta}_m^2 \gamma_n \\
\label{gap4}
&&{g_r \over (g_r)^2 + (h_r)^2} = \kr - {4 \over \beta} \sum_{l+s = -r} \hat{g}_s \Delta_l^2
- {4 \over \beta} \sum_{l+s = -r} \hat{\Delta}_l^2 g_s \\
\label{gap5}
&&{h_r \over (g_r)^2 + (h_r)^2} = - R + {4 \over \beta} \sum_{l+s = -r} \hat{\Delta}_l^2 h_s~.
\eea
By comparing (\ref{gap2}) and (\ref{gap3}), note that $\gamma_l = - R
\Delta_l^2$.

An important aid in finding numerical solutions to the gap equations
is to solve for the large-momentum behavior of the propagators
analytically \cite{KLL}.  At large momentum we find that the
propagators have the behavior
\bea
\nonumber
&& \Delta_l^2 \approx {1 \over (2 \pi l / \beta)^2 + m_\Delta^2} \qquad
\gamma_l \approx {- R \over (2 \pi l / \beta)^2 + m_\gamma^2} \qquad
\epsilon_l^2 \approx {(2 \pi l / \beta)^2 \over (2 \pi l / \beta)^2 + m_\Delta^2} \\
\label{asymptotics}
&& g_r \approx {2 \pi r / \beta \over (2 \pi r / \beta)^2 + m_\Delta^2} \qquad
h_r \approx {- R \over (2 \pi r / \beta)^2 + m_h^2}
\eea
where the asymptotic masses are given by
\bea
\label{AsymptoticMass}
m_\Delta^2 & = & R^2 + {4 \over \beta} \sum_l \hat{\Delta}_l^2 + {4 \over \beta} \sum_l \Delta_l^2 \\
\nonumber
m_\gamma^2 & = & R^2 + {4 \over \beta} \sum_l \hat{\Delta}_l^2 + {8 \over \beta} \sum_l \Delta_l^2 + {4 \over \beta R} \sum_l \gamma_l \\
\nonumber
m_h^2 & = & R^2 + {4 \over \beta} \sum_l \hat{\Delta}_l^2 + {8 \over \beta} \sum_l \Delta_l^2 + {4 \over \beta R} \sum_r h_r\,.
\eea

%%%%%%%%%%%%%%%%%%%%%%%%%%%%%%%%%%%%%%%%%%%%%%%%%%%%%%%%%%%%%%%%%%%%%%%%%%%%%%%%

\end{document}